*Review Article*

# Gyrotrons for High-Power Terahertz Science and Technology at FIR UF

**Toshitaka Idehara[1] and Svilen Petrov Sabchevski[2]**

[1]*Research Center for Development of Far-Infrared Region, University of Fukui, 910-8507 Fukui, Japan*
[2]*Institute of Electronics of the Bulgarian Academy of Sciences, 1784 Sofia, Bulgaria*

Correspondence should be addressed to idehara@fir.u-fukui.ac.jp

In this paper, we present the recent progress in the development of a series of gyrotrons at FIR UF that have opened the road to many novel applications in the high-power Terahertz science and technology. The current status of the research in this actively developing field is illustrated by the most representative examples in which the developed gyrotrons are used as powerful and frequency tunable sources of coherent radiation operating in a CW regime. Among them are high-precision spectroscopic techniques (most notably DNP-NMR, ESR, XDMR, and studies of the hyperfine splitting of the energy levels of positronium), treatment and characterization of advanced materials, new medical technologies.

## 1. Introduction

Till recently, and occasionally even today, the region of the electromagnetic spectrum that is situated between the microwaves and the far-infrared light has been considered as less exploited and has been termed figuratively a THz power gap. Nowadays, the remarkable progress in the development of both electronic and photonic devices (sources, waveguides and detectors) that operate in this frequency range (0.3-3.0 THz) is changing dramatically the situation [1, 2]. For example, in a quest for an ultimate performance the solid state devices (SSD) have reached unprecedented levels of energy efficiency and miniaturization. This makes them appropriate sources of radiation in many low-power THz technologies (e.g. sensing and inspection, imaging, short range communications). The output power of the SSD, however, is not sufficient for many applications that require power levels several orders of magnitude higher. Similar is the situation with the classical electron tubes (for instance slow-wave devices such as TWT, BWO and klystrons). Although at present they are experiencing a kind of renaissance and have demonstrated many outstanding achievements, their principle of operation (based on the utilization of tiny slow-wave structures with characteristic dimensions comparable to the wavelength of the radiation) limits their advancement towards THz frequencies [2, 3]. The free- electron devices (FEL, sources based on electron accelerators) are capable of delivering high peak output power in short pulses but are bulky and expensive. In comparison with all above-mentioned devices, the gyrotrons are the most powerful sources of coherent CW radiation in the sub-THz to THz frequency range [4]. As fast-wave devices that operate on a physical principle (known as electron cyclotron maser instability) that do not require slow wave structures, the gyrotrons can use much more powerful electron beams that interact effectively with the high-frequency electromagnetic field in oversized resonant structures [5, 6]. Therefore, the limitations of the slow wave devices are not characteristic for the gyrotrons and this stipulates their high-power capabilities.

One of the most prominent and already classic application of high-power (MW class) gyrotrons is the electron cyclotron resonance heating (ECRH) and current drive (ECCD) of a magnetically confined plasma in reactors (tokamaks and stellarators) for fusion [7]. Another well-known technology is the sintering of ceramic materials. Besides in these (already classic) processes the gyrotrons are being used in a number of novel and emerging research fields [8-13]. Among them are advanced spectroscopic techniques, electron cyclotron ion sources (ECIS), detection of concealed radioactive materials, beamed microwave propulsion and transmission, novel medical technologies, etc., just to name a few. The gyrotrons developed at FIR UF cover a broad field of output parameters (power and frequency) and are appropriate radiation sources for various studies and high-power THz technologies [12, 14]. In recent years, however, the main focus was on different advanced spectroscopic techniques (e.g. ESR, DNP-NMR, XDMR). For this reason, most of the tubes have been designed and optimized taking into account the specific requirements of any particular experimental set up for these methods.

In this survey, we present the recent progress in the development of a series of gyrotrons at FIR UF that have opened the road to many novel application in the high-power Terahertz science and technology. The current status of the research in this actively developing field is illustrated by the most representative examples in which the developed gyrotrons are used as powerful and frequency tunable sources of coherent radiation operating in a CW regime.

## 2. Gyrotrons of the FU CW Series

The first line of gyrotrons developed at FIR UF was the Gyrotron FU Series. It includes several tubes that are listed in Table I. As can be seen, they cover a wide frequency range from 38 GHz to 0.889 THz. The highest frequency of 0.889 GHz was reached by FU IVA, which utilizes a superconducting magnet with a maximum field intensity of 17 T and operates at the second harmonic of the cyclotron frequency. This achievement was a long-standing world record until it was exceeded by FU CW III (which will be presented below). These gyrotrons have been used as radiation sources for electron spin resonance (ESR) experiments (FU E and FU IVA), X-Ray Detected Magnetic Resonance, XDMR (FU II), and plasma scattering measurements (FU IA). Additionally, they have been used for systematic studies on the physics of gyrotron operation. For example, a series of experiments for investigation of modes interaction phenomena (such as competition, co-operation, switching) have been carried out with FU II. At the same time, other important processes, for instance, modulation of both the frequency and the output power and the operation at high harmonics have been studied in experiments using FU III. High mode purity and stable generation during long time periods have been demonstrated by FU V. It should be underlined, that all members of this series operate in a pulsed regime with a typical pulse duration of the order of several milliseconds, and a repetition rate of several Hz (i.e. with a duty ratio of several percent). For many applications, however, such pulsed operation is not appropriate. This motivated the development of a new series of radiation sources (FU CW) that provide a continuous wave (CW) radiation. Based on the experience gathered during the development, study and operation of the preceding series a number of new tubes have been designed and manufactured. The current members of the FU CW series are presented in Table II.

TABLE I: Gyrotrons of the FU Series

| Gyrotron | Frequency range, THz |
|---|---|
| FU I | 0.038-0.220 |
| FU E | 0.090-0.300 |
| FU IA | 0.038-0.215 |
| FU II | 0.070-0.402 |
| FU III | 0.100-0.636 |
| FU IV | 0.160-0.847 |
| FU IVA | 0.160-0.889 |
| FU V | 0.186-0.222 |
| FU VI | 0.064-0.137 |

TABLE II: Gyrotrons of the FU CW Series

| Gyrotron | Frequency range, THz | Max. output power, W | Max. B, T |
|---|---|---|---|
| FU CW I | 0.300 | 2.3x10$^3$ | 12 |
| FU CW II, IIA, IIB | 0.110–0.440 | 20–200 | 8 |
| FU CW III | 0.130–1.080 | 10–220 | 20 |
| FU CW IV | 0.134–0.139 | 5–60 | 10 |
| FU CW V | 0.2034 | 100–200 | 8 |
| FU CW VI, VIA, VIB | 0.393–0.396 | 50–100 | 15 |
| FU CW VII | 0.203–0.395 | 200 (50 CW) | 9.2 |
| FU CW VIIA | 0.132–0.395 | 200 | 8 |
| FU CW VIII | 0.100–0.350 | 100 | 8 |

The most powerful gyrotron of this series is FU CW I (see Fig.1), which delivers a CW Gaussian beam with a power of up to 2.3 kW at a frequency of 0.3 THz. It has been designed and manufactured by Gycom Ltd. as a radiation source for microwave processing (sintering) of ceramic materials. FU CW I has a He-free 12 T superconducting magnet and an internal mode converter that transforms the operating mode (TE$_{22,8}$ at the fundamental cyclotron frequency) in a well-collimated beam. The radiation is coupled by a corrugated waveguide and transmitted to the applicator, where the thermal treatment takes place using either a volumetric or surface heating, i.e. uniformly distributed (as a result of multiple reflections) microwaves or a focused wave beam, respectively.

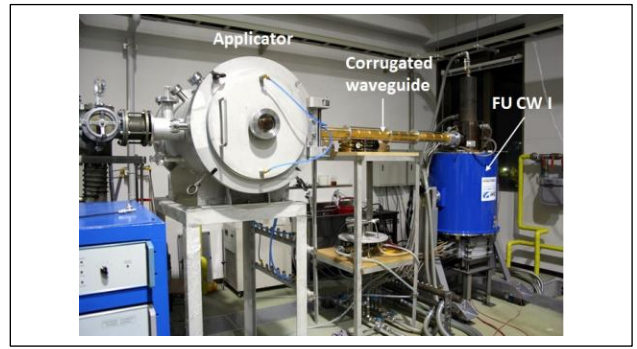

FIGURE 1: Gyrotron FU CW I equipped with a corrugated waveguide and an applicator

The gyrotron FU CW II is shown in Fig.2a. Its tube is of a demountable type and uses an 8 T cryo-free superconducting magnet and a triode magnetron injection gun. The resonant cavity has been designed for the second harmonic operation on $TE_{2,6}$ mode at 394.6 GHz. As it will be shown in the next section, its output parameters are appropriate for a radiation source for spectroscopic studies. An optimized variant of this device is FU CW IIA (Fig. 2b), which has a similar construction but instead of a demountable has a sealed-off tube.

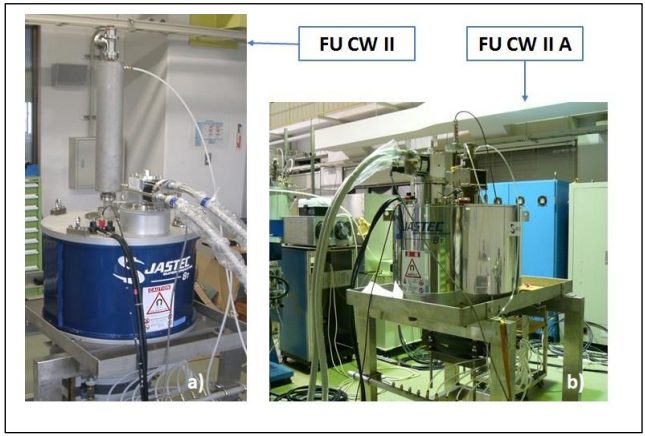

FIGURE 2: Gyrotrons FU CW II (a), and FU CW II A (b).

One of the most remarkable members of the FU CW series is the gyrotron FU CW III (Fig. 3a), which for the first time has surpassed the symbolic threshold of 1 THz and reached the highest frequency of 1.08 THz in a CW regime [15]. This breakthrough has been achieved at a second harmonic operation of the mode $TE_{4,12}$ using a 20 T superconducting magnet. The resonant cavity of this tube is identical with that of its predecessor – the first THz gyrotron with a pulsed ice-protected solenoid with a maximum field intensity of 20.5 T. It should be mentioned that almost all modes that have been exited in the pulsed tube have been excited in a CW operation as well. Moreover, FU CW III is step tunable in a wide frequency interval from 0.1 THz to 1.08 THz (see Fig 3b).

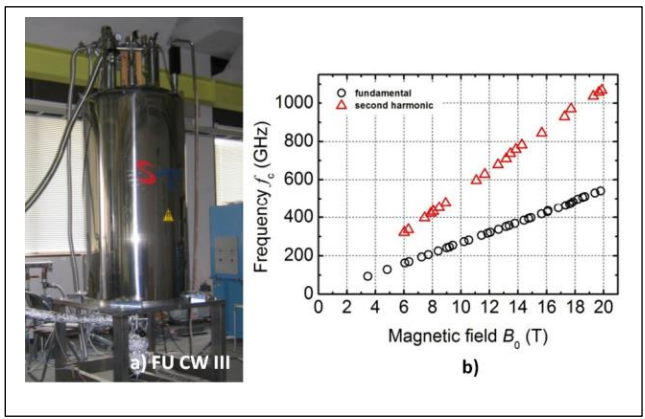

FIGURE 3: Gyrotron FU CW III (left) and frequency vs. magnetic field for the excited modes that demonstrate its step-tunability (right).

One of the gyrotrons in this series, which has been used as a versatile radiation source in the greatest number of various experiments, is FU CW IV (Fig.4a). This is due to its continuous frequency tunability, achieved by changing the magnetic field and exciting successively a sequence of high order axial modes (HOAM) $TE_{mnl}$, where the axial index takes the values $l = \pm 1, -2, -3, -4, ...$ (see Fig. 4b) [16].

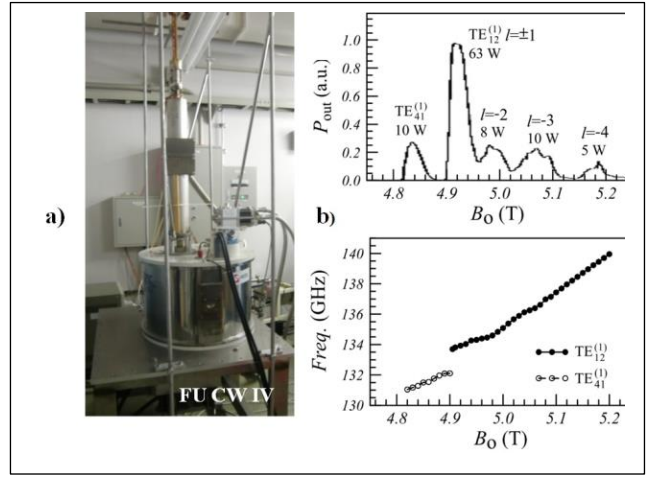

FIGURE 4: Gyrotron FU CW IV, (a) and its output parameters, (b) illustrating continuous frequency tunability from 134 GHz (at 4.9 T) to 140 GHz (at 5.2 T).

The next gyrotron, FU CW V has been designed for experiments dedicated to studies on the energy levels of positronium (Ps). It operates on the $TE_{03}$ mode (at the fundamental cyclotron resonance) and delivers radiation with a frequency around 0.203 THz and an output power of about 600 W [17]. A photo of this tube is shown in Fig. 5.

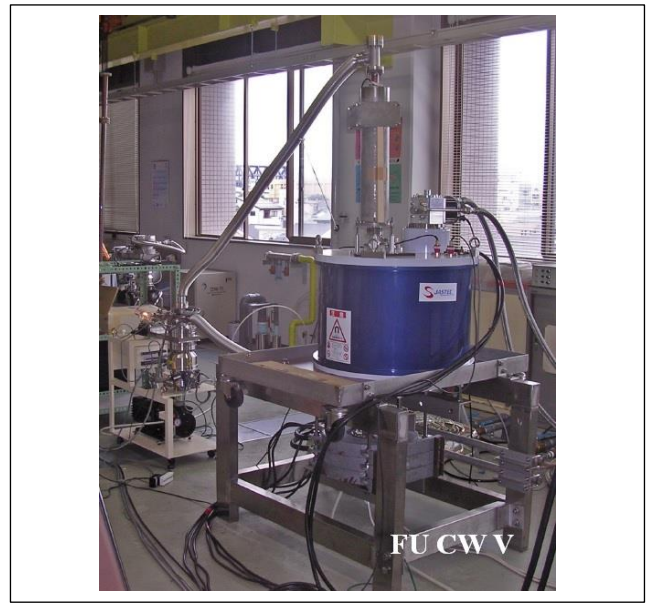

FIGURE 5: Gyrotron FU CW V with an 8 T superconducting magnet.

Another frequency tunable gyrotron is FU CW VI. Its design is based on a 15 T superconducting magnet. This tube also operates on a sequence of HOAM at the corresponding fundamental resonances of the cyclotron frequency using a backward wave interaction and exiting successively $TE_{06l}$ modes by controlling the magnetic field intensity in the cavity resonator. The frequency of the radiation can be tuned in a 2

GHz band from 394.6 GHz (at the lowest axial mode with l=1) up to 396.6 GHz. A photo of this device is shown in Fig.6. It has two versions (FU CW VIA and FU CW VIB) that have been specially optimized for the specific applications as radiation sources for DNP-NMR spectroscopy.

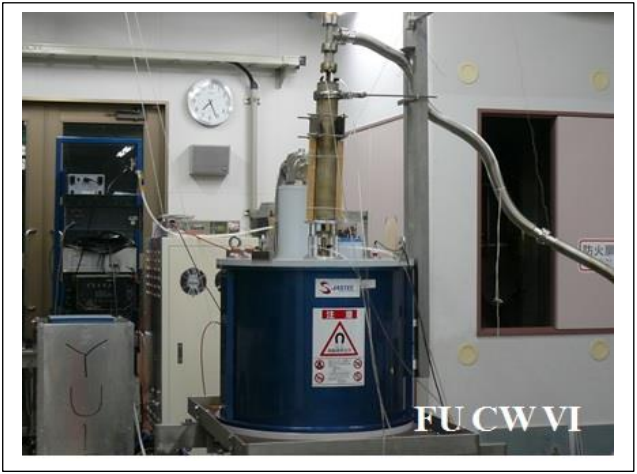

FIGURE 6: Gyrotron FU CW VI.

The gyrotron FU CW VII operates in the frequency range from 0.203 THz to 0.395 THz and delivers a radiation power of several hundred watts in a pulsed regime and about 50 W in CW. Distinguishing features of this oscillator are its excellent spectral characteristics that remain stable during long time periods. For example, the spectral width at half maximum (FWHM) is only several MHz. The fluctuations of the output power are less than 5%, wherein the frequency stability is better than $5.0 \times 10^{-6}$. FU CW VII is step-tunable from 86 GHz to 0.223 THz at the fundamental and from 0.187 to 0.430 THz at the second-harmonic resonances. A novel modified version (FU CW VIIA) has been developed recently for ESR spin-echo experiments that require high power and short pulses (see Fig. 7). For the experiments with FU CW VIIA an external quasi-optical mode converter and a millimeter-wave pulse forming system have been developed. The latter uses light controlled semiconductor shutters that consist of a pair of Si wafers. These nanosecond switches are being activated by high-energy Nd:YAG lasers. The system is capable of producing nanosecond millimeter-wave pulses with an arbitrary delay time of the order of microseconds.

The radiation source FU CW VIII is built using an 8 T superconducting magnet. It operates and is continuously tunable in a 2 GHz band around the central frequency of 0.203 THz. The output parameters of this gyrotrons are suitable for experiments on a precise measurement of the hyperfine splitting (HFS) of the positronium ground state.

It is evident from the photos of the gyrotrons presented above that their dimensions are determined mainly by the sizes of the used relatively bulky superconducting magnets. From the other side, it is clear that when used as radiation sources in a sophisticated laboratory environment it is easier to embed them into the rest of the equipment if their dimensions are smaller. In recent years, the availability of compact cryo-free superconducting magnets has made it possible to develop much more compact gyrotrons [18-20]. According to the nomenclature adopted at FIR UF, these recently developed devices are marked by the letter "C", which stands for "compact". Another recent extension of the FU CW series is a line of devices that have an internal mode convertor and radiate well-collimated Gaussian wave beams. They are marked by the letter "G". The current status of both "C" and "G" lines, i.e. FU CW C and FU CW G is presented in Table III.

TABLE III: Gyrotrons of the FU CW Series: C and G lines

| Gyrotron | Frequency range, THz | Max. output power, W | Max. B, T |
| --- | --- | --- | --- |
| FU CW CI | 0.395 | 120 | 8 |
| FU CW CII | 0.203 | $0.9 \times 10^3$ | 8 |
| FU CW GI<br>FU CW GIA | 0.203 | $0.5 \times 10^3$<br>$1.6 \times 10^3$ | 8 |
| FU CW GII | 0.2059<br>0.3934 | 110<br>74 | 8 |
| FU CW GIII | 0.3954 | 420 | 8 |
| FU CW GIV | 0.394–0.395 | 50 | 8 |
| FU CW GV | 0.162-0.265 | $1.3 \times 10^3$ | 10 |
| FU CW GO-I<br>FU CW GO-II | 0.460 | 100 | 8 |

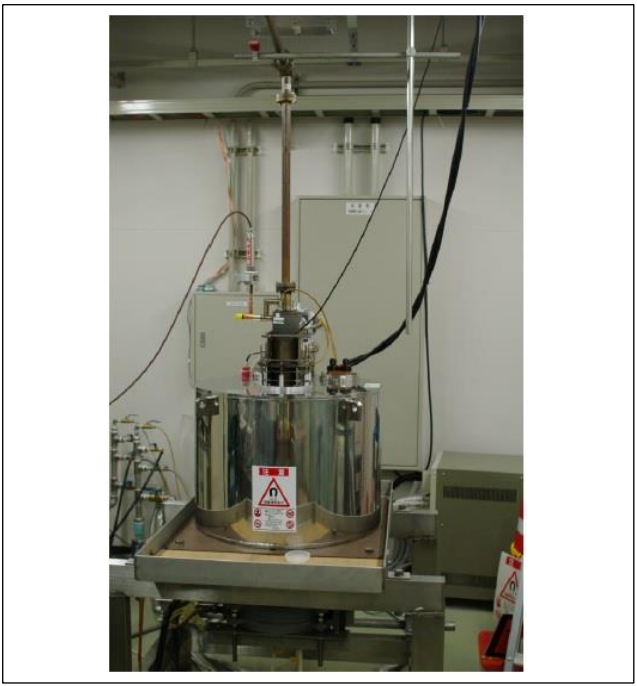

FIGURE 7: Gyrotron FU CW VIIA.

The compact gyrotrons FU CW CI and FU CW CII (see Fig 8 and Fig. 9) have smaller overall volume, mass and dimensions and can easily be integrated into the corresponding experimental setup. Their electron-optical systems (EOS) have specially designed magnetron injection guns (MIG) with minimized radial dimensions in order to accommodate them to the reduced diameter of the inner bore of the most compact magnets. For example, the tube of FU CW CI is only 1.02 m long and has a maximum diameter of 52 mm. The overall length of FU CW CII (0.86 m) with a Gaussian-like beam output is even smaller. While FU CW CI is a frequency tunable source in the range 0.107-0.205 THz and delivers radiation with an output power of 150-320 W, the next compact gyrotron (FU CW CII) operates at a fixed frequency of 0.203 THz at power levels about 0.8 kW.

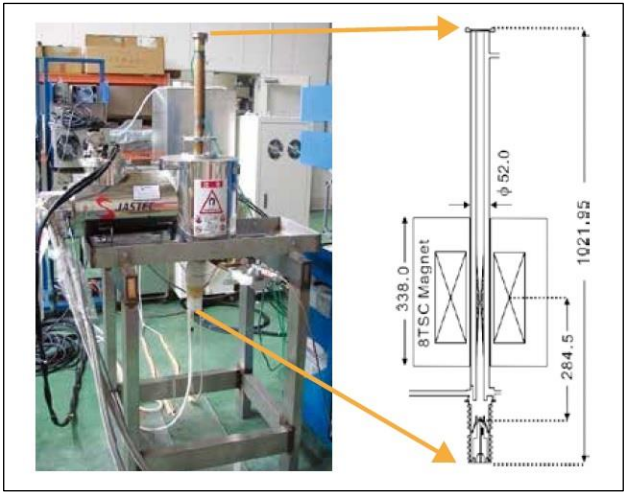

FIGURE 8: Compact gyrotron FU CW CI

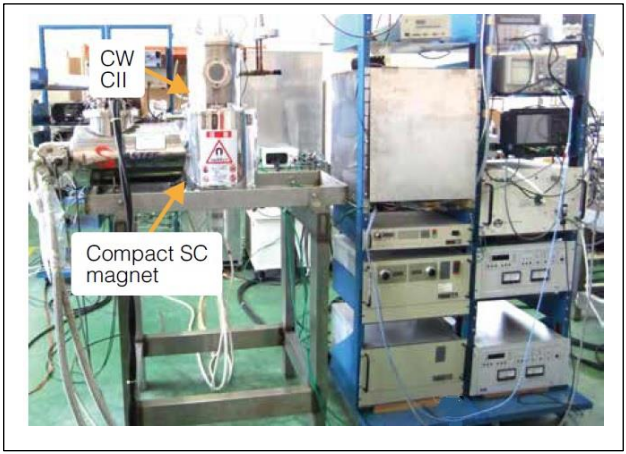

FIGURE 9: Compact gyrotron FU CW CII

The first compact gyrotrons with internal quasi-optical mode converters that have been developed at FIR UF are FU CW CII, FU CW GI and FU CW GII [21-23]. They consist of a helically-cut Vlasov type launcher and a set of four mirrors that form a Gaussian-like wave beam with a circular cross section. Fig. 10 shows the photo of FU CW GI, which has demonstrated a well collimated output beam with a power of 0.5 kW and a frequency of 202.55 GHz operating at the fundamental resonance of the counter-rotating $TE_{5,2}$ mode. The next tube with a Gaussian beam output is FU CW GII. It has an analogous design but a different resonant cavity, which has been optimized for an operation at the second-harmonic resonance.

The gyrotron FU CW GIII (see Fig. 11) has been designed and optimized for an operation at a frequency of 0.395 THz [24]. It radiates an output power of 0.4 kW with an efficiency of 4% at the second harmonic of the cyclotron resonance. A remarkable result demonstrated by this tube is an uninterruptable and stable CW operation for more than ten hours. The next device, FU CW IV [25] oscillates at the same frequency (second harmonic operation) but has an extended tunability band of almost 3 GHz.

The gyrotron FU CW V [26] is a multi-frequency gyrotron, which operates in the range from 0.162 to 0.265 THz at frequencies separated by steps of about 10 GHz. It has a double-disk output window with a variable spacing. The output power for almost all excited modes exceeds 1 kW.

A characteristic feature of FU CW VI [27] is a high-speed frequency modulation within a 100 MHz band. Its optimized version has demonstrated a frequency tunability in a band wider than 1.5 GHz. Both sources are build using a 10 T superconducting magnets and generate radiation (in CW and pulsed regimes) with a power of 20-30 W at frequencies close to 0.46 THz.

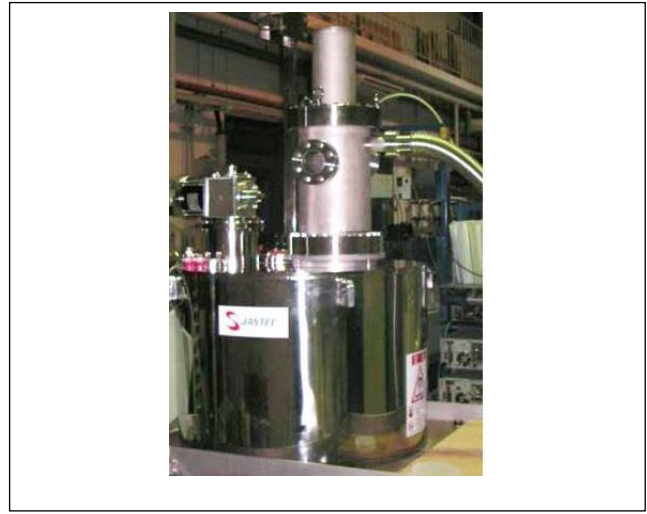

FIGURE 10: Compact gyrotron with a Gaussian beam FU CW GI

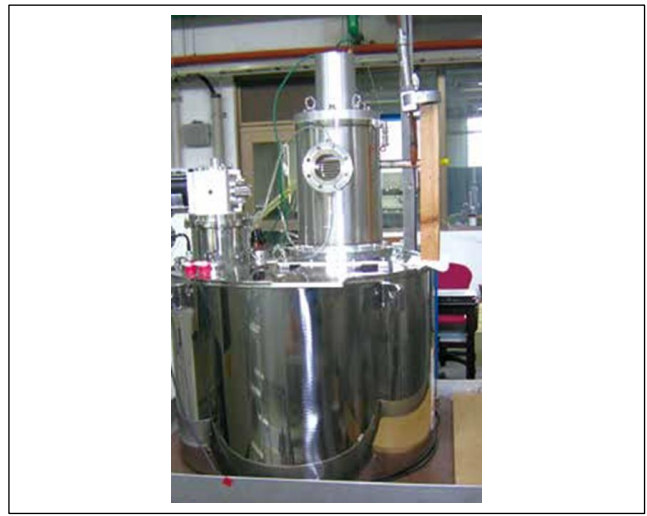

FIGURE 11: Compact gyrotron with a Gaussian beam FU CW GIII

# 3. Application of the FU CW gyrotrons as radiation sources to advanced spectroscopic methods

## 3.1 DNP-NMR spectroscopy

Nuclear magnetic resonance (NMR) spectroscopy is one of the most advanced methods for determination of the structures of both inorganic and organic molecules in material science and structural biology. At the same time, it has severe limitations due to its inherently low sensitivity and signal-to-noise (SN) ratio. A powerful technique for signal enhancement is the dynamic nuclear polarization (DNP). Its idea is to transfer the large electron spin polarization (usually using an chemical agent as electron polarization source) to the nuclear spin system irradiating the sample by microwaves with a frequency that is equal or close to the Larmor frequency of an electron (which is proportional to the intensity of the magnetic field in the spectrometer's magnet). Therefore, NMR spectroscopy with a signal enhancement through DNP is called DNP-NMR for short. For the first time, gyrotron radiation has been used for NMR-DNP in the pioneering experiments of R. Griffin at MIT. Since then, the gyrotrons have proved to be the most appropriate radiation sources, especially at high magnetic fields and, correspondingly, at high frequencies [28-34].

Some of the gyrotrons of the FU CW series have been developed especially for DNP-NMR and have been used in several spectrometers (see Table IV) at different institutions. Here, we outline briefly some of the most representative examples.

TABLE IV: DNP-NMR spectrometers and their radiation sources from the FU CW series of gyrotrons

|  | DNP-NMR 700 MHz | DNP-NMR 600 MHz | DNP-MNR 300 MHz | DNP-NMR 200 MHz |
|---|---|---|---|---|
| Radiation source | FU CW GO-I, GO-II | FU CW II, VI | FU CW VII | FU CW IV |
| Frequency, THz | 0.460 | 0.394 | 0.187 | 0.131 |
| Enhancement factor | ~ 23 | ~ 22 | ~ 60 | ~ 40 |
| Institution | Osaka Univ. | Osaka Univ. | Warwick Univ. | FIR UF |

The gyrotrons FU CW II, VI, VIA have been installed and are being used as radiation sources for the 600 MHz DNP-NMR spectrometer of the Institute for Protein Research at Osaka University. According to the nomenclature adopted by the research team there, FU CW VI and VIA are named FU CW GO-I and GO-II, where "GO" indicates that the tube delivers an optimized Gaussian output beam. Their photos, taken at Osaka University, are shown in Fig. 12. Recently, these devices equipped with a transmission line (supplied by Bridge 12 Co., Ltd.) have been installed in the novel 700 MHz system [35]. The quasi-optical transmission system combines two independent sub-millimeter waves produced by this couple of 0.46 THz gyrotrons into a single dichromic wave. An advantageous feature of the irradiation system is that the used gyrotrons are not only frequency tunable but can deliver also frequency modulated radiation [27, 36]. This additional function is beneficial for increasing further the enhancement of the signal. Other important factors are the stability and the precise control of the output parameters of the gyrotron. A proportional-integral-derivative (PID) feedback control system has been developed, which decreases the beam current fluctuations from 17 % to 0.5% and these of the output power from 15 % to 5 % by controlling the current of the heating filament of the cathode. More effective stabilization of the output power (up to 2%) has been realized by a feedback control that adjusts the voltage of the heater [37-39].

In the first study of high-field DNP with a magnetic field intensity higher than 10 T, carried out at Osaka University, a tenfold increase of the signal has been observed for a $^{13}$C-labeled organic compound. The frequency tunability of the used gyrotrons over a range of more than 1 GHz has allowed optimizing the DNP resonance condition without a time-consuming sweeping of the field of the NMR magnet. As a result, a significant enhancement of the solid-state NMR signal at 90 K has been achieved. In a series of experiments the dependence of the enhancement on the temperature of the sample has been studied in detail under both static and MAS (magic angle spinning) conditions at a high external field strength of 14.1 T in the range from 30 K to 90 K. It has been observed that in the case of static DNP the cooling of the sample from 90 K to 35 K leads to a doubling of the enhancement (from 11 to 23). An additional two- to three-fold increase of the enhancement has been obtained with the sample spinning (MAS). The maximum DNP enhancement of 50 has been achieved under MAS at a temperature of 33 K. The total sensitivity gain at 30 K and magnetic field of 14.1 T (≈780) is more than twice that obtained at 90 K and 9 T (≈400), and three times that at 90 K and 5 T (≈240) [40].

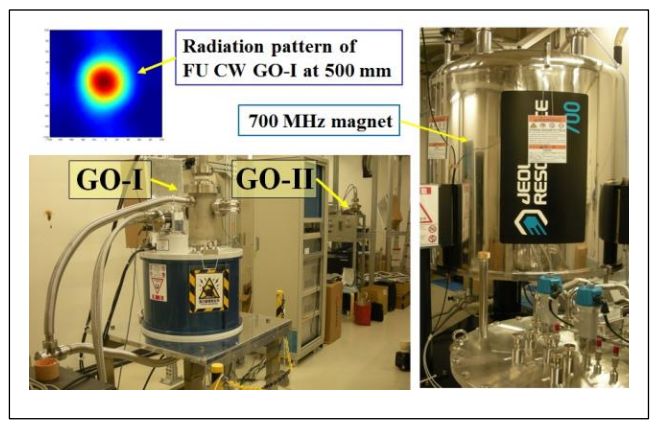

FIGURE 12: Gyrotrons FU CW GO-I and GO-II (left) installed in the 700 MHz DNP-NMR spectrometer at Osaka University.

FU CW VII, can be used for both 300- and 600 MHz DNP-NMR at output frequencies of 200 GHz (fundamental operation) and 400 GHz (second harmonic of the cyclotron resonance), respectively. This versatile tube, which is step-tunable in a wide range (from 86 to 223 GHz) has been installed in the 300 MHz spectrometer (see Fig. 13) of the NMR Group in the Department of Physics at Warwick University [41]. In this system, the radiation of the gyrotron is converted into a Gaussian beam and then transmitted to the NMR probe by an optical bench, which has beam splitters for monitoring and controlling of the microwave power. Additionally, it uses a ferrite rotator for stopping the reflected radiation, and a Martin-Puplett interferometer for adjusting the polarization. Through a corrugated waveguide the wave is coupled to the MAS NMR probe. The enhancements that have been observed at 6.7 T and a microwave pulsed power of only 1 W are as follows: (i) $^{13}$C signal enhancement of 60 for a frozen urea solution; (ii) 16 for transmembrane protein bacteriorhodopsin (bR) in two-dimensional crystalline patches (aka "purple membrane"); and (iii) 22 for $^{15}$N in a frozen glycine solution. The efficiency of the enhancements can be illustrated by the following estimates. Even a modest enhancement of 16 (at 6.7 T and only about 1 W microwave power) reduces the necessary experimental time by a factor of 256. At 90 K, both the corresponding Boltzmann factor and the reduced thermal noise provide a further gain of about 4 compared with a room temperature operation. In total, this gives a potential total reduction of the experimental time of ≈4000 times [41].

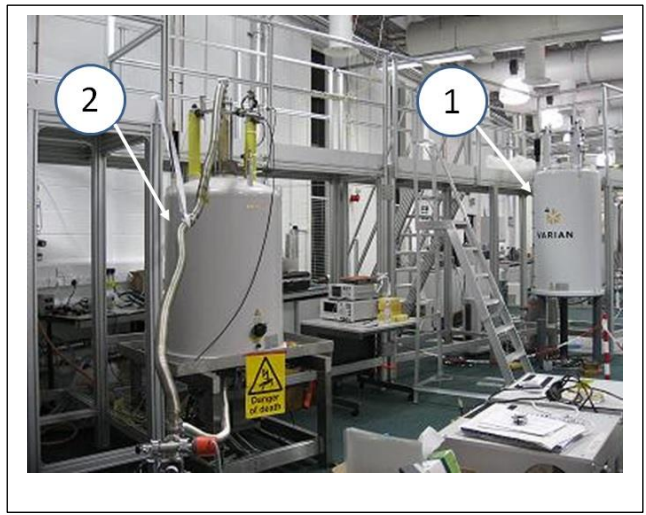

FIGURE 13: DNP-NMR 300 MHz spectrometer at Warwick University, UK (1) utilizing the gyrotron FU CW VII (2) as a radiation source.

In FIR FU, the gyrotron FU CW IV has being used as a radiation source for DNP-NMR studies at 200 MHz [42]. Fig. 14 shows the developed DNP-enhanced cross-polarization/magic angle spinning (DNP/CP/MAS) NMR system. It includes a 200 MHz Chemagnetics CMX-200 spectrometer with a 4.7 T magnet. The maximum enhancement of more than 30 has been obtained for C1 $^{13}$C-enriched D-glucose dissolved in a frozen medium containing mono-radical 4-amino-TEMPO. Using the same equipment the first spectra of PMMA submicron size particles have been obtained and analyzed. The results indicate that DNP/CP/MAS NMR can be used for characterization of the surface structure of nanomaterials (including polymeric materials).

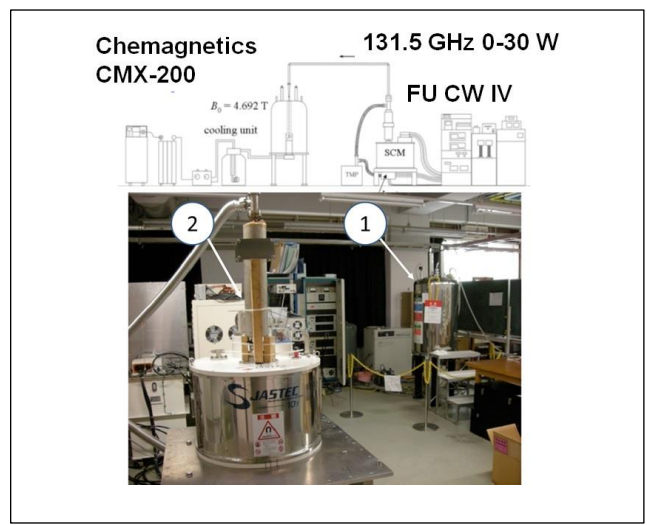

FIGURE 14: DNP-NMR 200 MHz Chemagnetics CMX-200 spectrometer at FIR UF (1) with the gyrotron FU CW IV (2) as a radiation source.

### 3.2 Precise measurement of the HFS of positronium

Positronium (Ps) is a bound state of an electron and a positron, which forms a Hydrogen like exotic atom that can exist in two states, namely ortho- (o-Ps) and para-positronium (p-Ps), respectively. The former is a singlet, which decays into two γ quanta, while the latter is a triplet that decays in three quanta. Due to its simplicity, Ps is an excellent object for studies on QED and particle physics. The splitting between the ground energy levels of o-Ps and p-Ps known as HFS (hyperfine splitting) is about 0.84 meV, which corresponds to a frequency of 203 GHz. The experimental values of the HFS obtained using indirect methods (e.g. Zeeman splitting in a static magnetic field), however, do not agree well with the theoretical predictions. The observed significant discrepancy of 3.9σ (15 ppm) has motivated the development of a novel and precise technique for evaluation of the HFS [43]. It is based on a stimulated transition from o-Ps to p-Ps induced by means of irradiation by a strong electromagnetic wave with a frequency of about 203 GHz. The estimates show that the gyrotrons are the only CW sources in this frequency range that can provide the necessary radiation power. Therefore, in order to make possible the realization of the novel direct method for precise measurements of the HFS several gyrotron tubes (most notably FU CW V) with appropriate output parameters have been designed and manufactured in FIR UF. The collage in Fig. 15 presents schematically the used experimental set up. Schematically the experimental setup is shown in Fig. 15. It consists of the gyrotron FU CW V as a radiation source, a quasi-optical mode converter and transmission line that includes a Vlassov type launcher (L) and a set of four mirrors (M0-M3), Fabry-Perot (FP) cavity, $^{22}$Na source of positrons, and two gamma ray detectors. The positronium is formed in the cavity using nitrogen as a stopping target. Using this experimental system the first direct measurement of the hyperfine transition of the ground state positronium has been carried out. Such transition has been observed with a significance of 5.4 standard deviations. The transition probability that has been measured directly for the first time is in a good agreement with the theoretical value [44]. The whole Breit-Wigner resonance of the transition from o-Ps to p-Ps has also been measured for the first time tuning the frequency of the gyrotron radiation in a very wide interval from 201 to 205 GHz by exchanging successively several gyrotron cavities of different radii [45]. As reported in [45], the developed system opens a new era of millimeter-wave spectroscopy, and enables to directly determine both the hyperfine interval and the decay width of para-Ps (p-Ps).

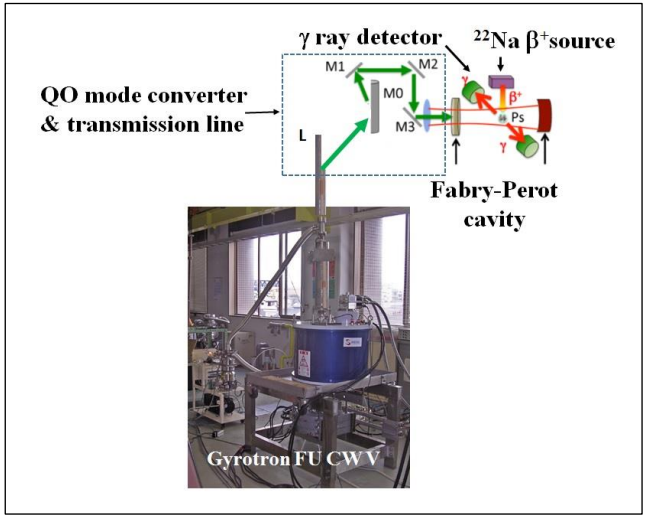

FIGURE 15: Experimental setup for a precise measurement of the HFS of positronium.

### 3.3 XDMR spectroscopy

The spectroscopy based on X-ray detected magnetic resonance (XDMR) is a novel pump and probe technique in which X-ray magnetic circular dichroism (XMCD) is used in order to probe the resonant precession of either spin or orbital magnetization components pumped by the magnetic field of a strong sub-THz gyrotron radiation in a plane perpendicular to the applied static magnetic field. The most prominent advantage of this technique is that it is both element- and edge-selective method that can be used for detailed studies of the precession dynamics of orbital and spin magnetization components. In order to prove the feasibility of this novel spectroscopic technique the gyrotron FU II has been installed at the beam line ID12 of the European Synchrotron Radiation Facility (ESRF) in Grenoble, France. Recent studies [46] have indicated that the same approach can be used also for investigation of various new X-ray electro-optical and magneto-electric effects.

The results of the experiments carried out so far indicate some possibilities for further development of this promising novel spectroscopic technique. For example, it has been suggested to extend the measurements to frequencies of up to 140 GHz in order to be able to investigate the dynamics of the Van Vleck orbital paramagnetism, to detect the high-frequency normal modes associated with magnetically coupled sublatices, etc. [46].

### 3.4 ESR spectroscopy

In FIR FU, the first experiments on a high-frequency electron spin resonance (ESR), a.k.a. electron paramagnetic resonance (EPR) spectroscopy have been carried out using as sources of microwave radiation the gyrotrons from the previous series (Gyrotron FU) [47]. For these experiments, an advanced ESR spectrometer with a pulse magnet with a maximum intensity of 40 T has been developed. It has been used for an investigation of the magnetic properties of various materials such as Fe-$SiO_2$ granular films, $CsFeCl_3$, powder spectra of $SrCu_2(PO_4)_2$ and $BaCu_2(PO_4)_2$ compounds and anti-ferromagnetic single crystal $MNF_4$.

Similar to the NMR, the ESR spectroscopy becomes more powerful at high magnetic fields and frequencies. Additionally, it benefits from an excitation by coherent pulses rather than continuous waves (see for example [48]). In recent years, such considerations have motivated the development of radiation sources and instrumentation for a pulse ESR, also referred to as Fourier Transform ESR (FT-ESR), in FIR UF [49]. This technique is characterized by a significant gain of the sensitivity and reduction of the spectrum acquisition time. The most widely used method is electron spin echo envelope modulation (ESEM). In order to realize this and other similar techniques a novel spectrometer has been developed. It uses as a radiation source the gyrotron FU CW VII and a millimeter-wave pulse forming system that have been described in Sec. 2. After the successful tests that have demonstrated the capabilities of the developed system, currently several experiments on spin echo ESR are in progress [50].

## 4. Other applications of the gyrotrons developed at FIR UF

### 4.1 Plasma diagnostics

In 1982, 70 GHz radiation from Gyrotron FU I was used for plasma scattering measurement on drift waves excited spontaneously in the W-II tokamak at Kyoto University. The pioneering experimental results are presented in Fig. 16. This is the first trial of the application of gyrotron for plasma diagnostics [51].

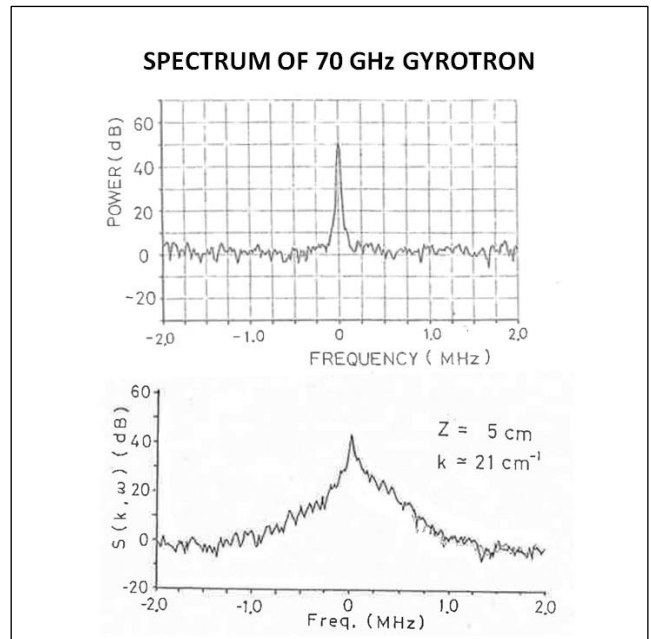

FIGURE 16: Frequency spectrum of the output radiation of the Gyrotron FU I (upper panel) and of the scattered signal (lower panel).

The frequency spectrum was measured by a heterodyne detection system, which includes a harmonic mixer that generates the intermediate frequency (IF) signal. On the spectrum, a mono-frequency signal is seen with a narrow width (upper panel of Fig.16). The broadening of the spectrum was observed when a lower hybrid wave heating (LHWH) was applied (lower panel). The width of the broadening was from several hundred kHz to 1 MHz. This means a low frequency. The plasma wave is excited spontaneously as an unstable wave during the LHWH. The drift wave instability is sometimes dangerous for the stable confinement because it leads to disturbance of the plasma. Therefore, the scattering measurement of drift waves is very important to achieve stable confinement of the plasma during LHWH.

As already mentioned, FU II is one of the most versatile tubes of the Gyrotron FU series since it has demonstrated many remarkable capabilities, which are appropriate for various experiments and, among them, for plasma diagnostics. It has been used successfully for scattering measurements in the studies on spontaneously excited drift waves during the Neutral Beam Injection (NBI) heating phase in the Compact Helical System (CHS) at the National Institute of Fusion Science (NIFS) in Japan. The experimental setup is shown in Fig. 17. It consists of a homodyne detection system and the gyrotron FU II as a radiation source.

In these experiments, the output radiation of the gyrotron is injected into the plasma and the scattered signal is observed by the homodyne detection system. The operation parameters of the gyrotron were as follows: the frequency and the output power of the gyrotron were 354 GHz and 110 W in pulsed regime at the second harmonic of the electron cyclotron frequency; the cavity mode was $TE_{161}$; and the pulse width and repetition rate were several hundred millisecond and typically several tens Hz, respectively. The measurement results are shown in Fig.18. The plasma is generated by electron cyclotron heating (ECH) and then it is heated further by NBIH. The pulse width of NBIH is 100 msec. The trace shows a series of frequency spectra observed from the intermediate frequency signal.

As shown in Fig. 18, during the whole time, the output radiation from the gyrotron is injected into the plasma and the homodyne detection system is in operation. In each observed spectrum, the frequency is swept from zero to 1 MHz with a sweeping time of around 20 msec. Therefore, the observed frequency spectra represent the time evolution of the spectra of the low-frequency fluctuations. It is seen well that the frequency broadening due to the excitation of low-frequency density fluctuation appears only during the NBI phase. This means that any instability occurs during NBI and low-frequency plasma wave (for example, Drift Wave) is spontaneously excited in the plasma. This is a kind of scattering measurement of plasma waves during an unstable phase of plasma.

The collective Thomson scattering (CTS) is an important technique for a diagnostic of magnetically confined plasma in various fusion reactors. It allows studying the distributions of thermal and fast ions with a high spatial resolution. Recently, several novel gyrotrons with output parameters that are appropriate for Collective Thomson Scattering (CTS) diagnostic in the Large Helical Device (LHD) at NIFS have been developed and used in a series of experiments [52, 53]. One of them operates at 0.398 THz (second-harmonic) and delivers a power of 83 kW. The other tubes provide radiation with frequencies of 0.295 THz and 77 GHz at output power levels of 0.22 MW and 1.9 MW, respectively [53]. At such elevated powers, the signal to noise ratio (S/N) and the overall efficiency of this plasma diagnostic technique increases significantly.

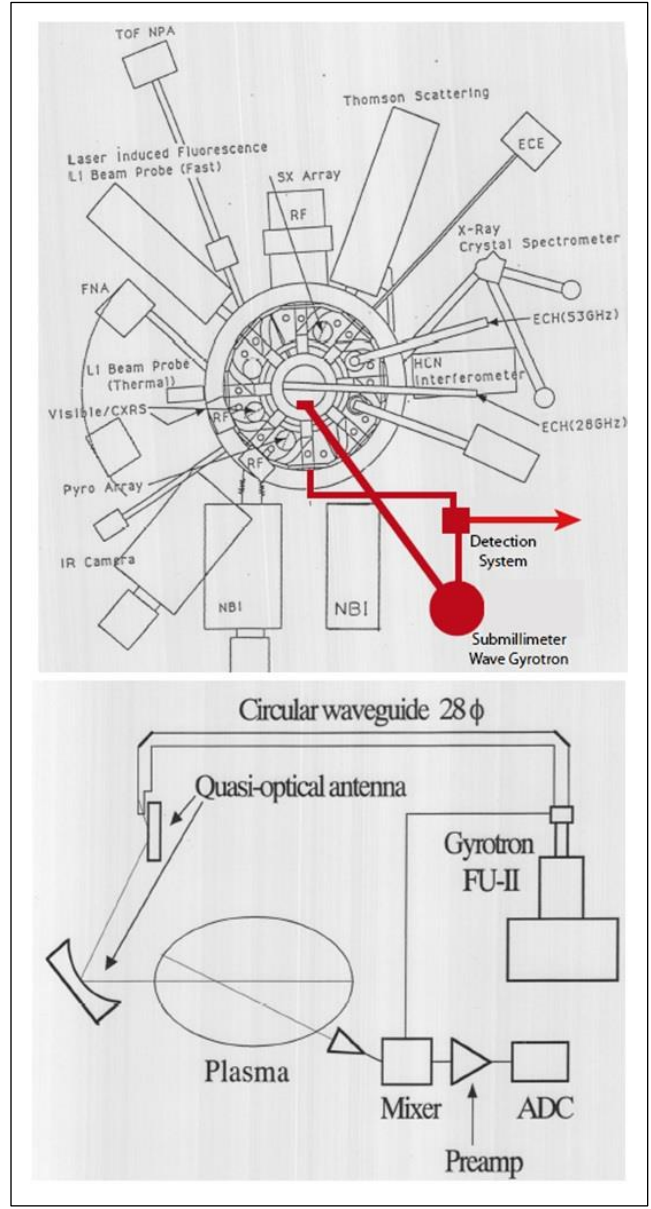

FIGURE 17: The installation of Gyrotron FU II and the homodyne detection system on CHS device at NIFS (upper panel), and the block diagram of the homodyne detection system (lower).

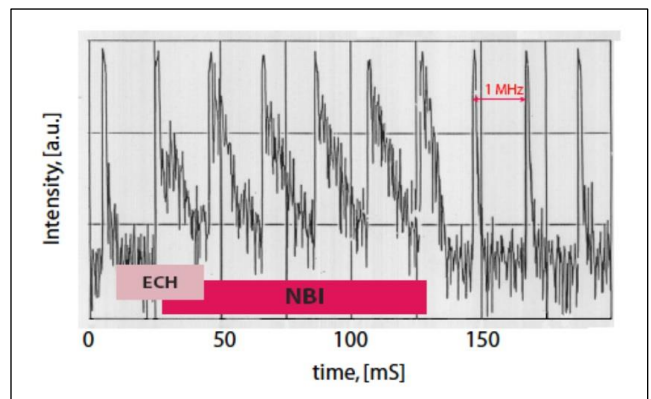

FIGURE 18: The measurement results obtained in CHS using a homodyne detection system. The time evolution of frequency spectra of the scattered signal from the spontaneously excited low frequency plasma waves (e.g. "Drift Wave") shows the frequency broadening of the observed spectra.

## 4.2 Processing of advanced materials

The thermal treatment of advanced materials (e.g. ceramics sintering) by microwaves has many well-known advantages compared with the traditional methods. For example, it provides more uniform and volumetric heating, better control of the heating rate, low thermal losses due to the contactless nature of the process, etc. The most frequently used sources of microwaves in the industrial technologies are the magnetrons. Their radiation, however, has a wavelength of the order of centimeters (for instance, about 12 cm at 2.45 GHz), which makes it difficult to achieve a homogeneous heating. In this respect, the gyrotrons have a great advantage since they can generate radiation at millimeter and sub-millimeter wavelengths. Moreover, at a shorter wavelength, such problems as the formation of hot-spots and thermal run away are considerably reduced.

At FIR UF, there are two versatile materials processing sysems that are currently in operation. They have been developed in a collaboration with IAP-RAS (Russia). The first one utilizes a 28 GHz gyrotron with a total power of 15 kW. The second one is a sub-millimeter wave system, which is built using as a radiation source FU CW I (0.3 THz, 2.5 kW). It is equipped with a large volume applicator in which the samples can be heated by a focused beam (surface treatment) using a focusing mirror or by uniform irradiation (in the case of volumetric heating) using a corrugated sheet and mode steerers mounted inside. Additionally, the applicator has six ports for measurements, two vacuum ports, and a peephole for visual observation.

On this equipment, different materials have been treated. Among them were oxide ceramics (alumina, silica), boron carbide ($B_4C$), zirconia and silica xerogel ceramics [54-58]. The obtained experimental results give a deeper insight into the underlying processes of the microwave heating and provide information about the influence of the technological parameters on the structure (grain size, porosity, etc.) and the physical properties (e.g. density, mechanical strength and so on). Therefore, they are useful for a further optimization of both the experimental parameters and the used equipment. In a number of experiments important and interesting from both theoretical and practical point of view non-thermal effects of the microwave processing have been observed and studied [59].

## 4.3 Novel medical technologies

For the first time in the world, experiments on irradiation of cancer tumors in animal's tissue have started at FIR UF many years ago using the gyrotrons of the previous series [60, 61]. Lately, these studies are being carried out more actively because the new series (FU CW) provide more convenient radiation sources for clinical trials. Fig. 19 shows the experimental setup used for i*n vivo* treatment of tumors implanted in experimental animals. In it, the radiation of the gyrotron FU CW IV is transmitted through a waveguide to the movable stage on which the animal is fixed. The temperature of the tumor is measured by thermocouples and the deposited energy is controlled by tuning the output power of the gyrotron and by adjusting the distance between the waveguide taper (which serves as an antenna applicator) and the stage. The results of the hyperthermia therapy carried out at frequencies of 0.107 THz and 0.203 THz are impressive and very promising. Even a stronger effect has been obtained combining the hyperthermia with a photodynamic therapy (PDT) and using a novel multifunctional photosensitizer [62].

Recently, both a concept and a dedicated radiation source (FU CW CI) for another hybrid therapy have been developed [63]. It is based on the simultaneous and/or sequential application of two beams, namely a beam of neutrons and a CW (continuous wave) or intermittent sub-terahertz wave beam produced by a gyrotron for a treatment of cancerous tumors. Schematically, the dual-beam irradiation facility is shown in Fig. 20.

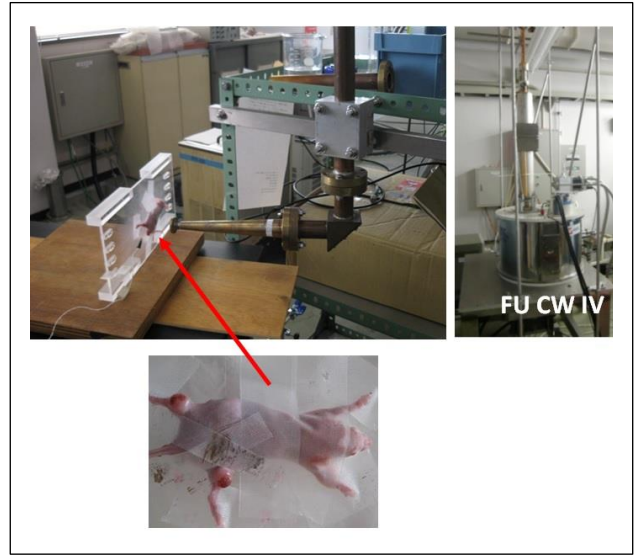

FIGURE 19: Experimental setup for *in vivo* irradiation of animals with cancerous tumors by sub-THz radiation: radiation source FU CW IV (right panel) and the irradiation stage (left panel).

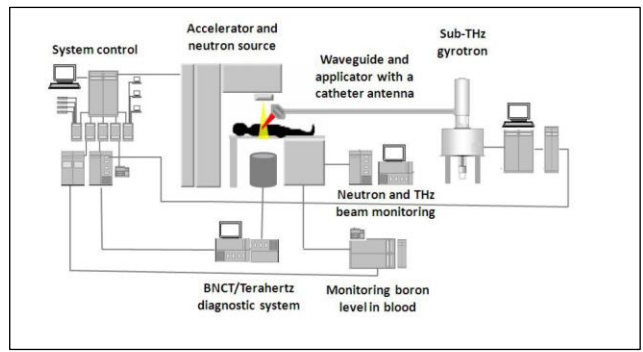

FIGURE 20: Schematic layout of the dual-beam irradiation system.

It consists of the following main components: (i) an accelerator based neutron source; (ii) a CW sub-THz gyrotron; (iii) a system for delivering the radiation to the treated zone of the patient's body which includes a waveguide transmission system and a catheter antenna; (iv) systems for monitoring and control of the parameters of both the neutron and the terahertz beams; (v) a diagnostic system for monitoring and control of the doses of irradiation; (vi) a system for monitoring of the boron level in the blood of the patient together with other biochemical parameters.

## 5. Conclusions

The gyrotrons of the FU CW series, developed at FIR UF have opened the road to various novel and prospective applications in many research fields. In order to satisfy the growing demand of CW radiation sources operating in the THz frequency range for various physical studies and technologies, this series is being extended continuously. Based on the accumulated experience, nowadays the development of new devices is notably accelerated. Moreover, the new members of the FU CW series are characterized by an improved performance and optimized output parameters according to the requirements imposed by each specific application. Therefore, we expect that this series of radiation sources will continue to grow both quantitatively and qualitatively.

It should be mentioned that similar studies are being carried out in many other institutions worldwide [4, 13, 64, 65]. The significant progress in the development of frequency tunable gyrotrons [28-31, 64, 65] is a driving force for further improvements of several advanced spectroscopic techniques, most notably NMR-DNP [69, 71]. Other active research fields that benefit from gyrotrons as powerful sources of THz radiation are plasma physics [72-77], remote detection of concealed radioactive materials [78-79], imaging and inspection [80], just to name a few.

It is well recognized that the international collaboration between the research groups working on the development and application of gyrotrons has always been very stimulating and fruitful. Recently, FIR UF has organized an international consortium for Development of High-Power Terahertz Science and Technology [81]. In this organization participate 13 institutions from 9 countries around the world. We believe that it will facilitate and boost further an active co-operation in this rapidly advancing research field.

# Acknowledgement

This work was supported partially by a Grant in Aid for Scientific Research (No. 25630142) from Japan Society for Promotion of Science (JSPS) and SENTAN Project of JSPS,